\documentclass[12pt]{article}
\usepackage[english]{babel}

\usepackage{amsmath}
\usepackage{graphicx}
\usepackage[colorlinks=true, allcolors=blue]{hyperref}

\renewcommand{\P}{\mbox{Pr}}
\newcommand{\SID}{\mbox{SID}}
\newcommand{\SND}{\mbox{SND}}

\usepackage{amsmath}
\usepackage{graphicx}
\usepackage[colorlinks=true, allcolors=blue]{hyperref}

\title{Lottery paradox, DNA evidence and other stories: How to accept uncertain statements }
\author{Yudi Pawitan\\
Department of Medical Epidemiology and Biostatistics\\
Karolinska Institutet\\
yudi.pawitan@ki.se}

\begin{document}
\maketitle

\begin{abstract}
I think we can agree that dealing with uncertainty is not easy. Probability is the main tool for dealing with uncertainty, and we know there are many probability-related puzzles and paradoxes. Here I describe a rather idiosyncratic selection that highlights the problem of accepting uncertain statements. Without going into a formal decision theory, there are simple intuitive rational bases for doing that, for instance based on high probability alone. The lottery paradox shows the logical problem of accepting uncertain statements based on high probability. The DNA evidence story is an example of the use probabilistic reasoning in court,  where philosophical differences between the schools of inference -- the frequentist, Bayesian and likelihood schools -- lead to substantial differences in the quantification of evidence.
\end{abstract}

\section{Lottery paradox}
Since we all face uncertainty all the time, one would think that we should be good at dealing with it. But we are not. Why is that? First, consider Kyburg's (1961) rule of acceptance: it is rational to accept an uncertain proposition that has a high probability of being correct. This sounds reasonable and one might argue we cannot live without this rule; we go about our daily life and make plans without thinking about various risks that we know are constantly around us, as we cross the road, drive to work, or take a plane journey, etc. We know accidents do happen, but with small enough probabilities that we presume they will not happen to us. In fact your family and friends might worry about you if you start worrying obsessively about minor risks in life.

Yet Kyburg came up with the lottery paradox to highlight the problem of accepting uncertain propositions, even those with very high probability. There is a large literature surrounding the paradox; see the Wikipedia entry to get an overview. Suppose you buy a ticket in a lotto where your chance of winning is 1 in a million. It is rational to presume you're going to lose and to go on living the same way as if you have not bought the lotto; it is irrational to make real plans -- like buying a fancy car or a big house -- on the off chance of winning the lotto. You would also expect every other lotto buyer to think the same way. Yet, invariably, there is a lotto winner. So, in effect, you're holding a contradiction: each buyer is going to lose, but one (or even more) buyer is going to win. 

A formal analysis of Kyburg's paradox is simple if we presume the lotto is a raffle, so each ticket number is sold once and there is a guarantee of one winner. Assuming, there are 100 tickets sold, the probability that a particular ticket $i$ loses is
$$
\P(\mbox{Ticket $i$ will lose})= 99/100,
$$
which is large enough to justify the feeling that you're going to lose. But, for a collection of $k$ tickets, the probability that all of them will lose is
$$
\P(\mbox{$k$ tickets will lose})= (100-k)/100,\  \mbox{for } k=1,\ldots,100,
$$
which drops linearly to zero as $k$ increases. When all tickets are sold, a winner is guaranteed.

Here is a feature that makes acceptance of uncertain statements difficult or non-intuitive. Certain/sure statements are governed by the more intuitive but rigid laws of logic: for instance, if statements $S_1$ and $S_2$ are individually true then the joint statement $S_1\land S_2$ must be true. We can combine any number of statements with no degradation of certainty in the conjunction of the statements. That is not the case with uncertain statements: conjunction will generally degrade the overall quality. In the lotto example, each statement that ticket $i$ will lose has a probability of 0.99; but the probability that two  tickets will jointly lose is degraded to $98/100$. The probability continues to degrade with larger and larger number joint statements until it goes to zero altogether. Our willingness to hold the contradictory positions --  each will lose but somebody will win -- is our concession to uncertainty; it's something we would or should be embarrassed to do when dealing with certainty.

That willingness might represent our implicit acceptance that our brain is not wired to think probabilistically, as it is easy to get lost in the conjunctive tangles of a joint event. In the so-called birthday paradox, first imagine two people comparing their birthdays. The probability they share a birthday is 1/365. So it seems reasonable to presume they will not celebrate their birthday together. When a third person comes, the probability of any pair of them to share a birthday is also 1/365, and so on for more and more people. But, jointly, considering all the pairs together, the chance of some people sharing birthdays will increase, but by how much? For instance, what happens in a room of, say, 25 people? Is it likely for two or more people to share a birthday? It turns out to be quite likely: the probability has increased to 0.57. In a room of 50 people, it is almost certain (probability 0.97) that some people can celebrate their birthdays together. Most people will find this unintuitive. 

\section{Probability in court}
Uncertainty rules in the court rooms, but at some point judges or juries have to accept and assimilate uncertain statements  provided by the prosecution and the defence. What acceptance rules do they follow? The basis of conviction in criminal cases is `beyond reasonable doubt.' What does that mean? It should be that the evidence provided is sufficient to banish any lingering doubt. Does that mean a level of probability of guilt so high that any remaining doubt is unreasonable? In civil cases, the standard of proof is on 'the balance of probability.' This sounds like a probability greater than 0.5, but in any case it definitely sounds less demanding than `beyond reasonable doubt.' As shown in the case of O.J.\ Simpson's murder trial, an acquittal in a criminal court may be followed by a loss in a civil court. 

The use of probability in court raises inherently difficult issues directly connected to the meaning of probability. First, let's consider probability as a long-term frequency: how does it apply to the case at hand? For example, suppose it is known (statistically) in a nameless corner of the world that 60\% self-employed businessmen cheat on their taxes. Mr.\ John is a self-employed businessman. On the `balance of probability' he is a tax cheater; but no judge will convict him on the basis of the statistical probability alone. If the statistical probability is sufficient for conviction, then every self-employed businessman should be convicted immediately. In the frequentist philosophy, the 60\% probability does not apply to Mr.\ John. His defence lawyer can easily produce all sorts of facts as evidence of his good characters, e.g., he is a faithful church-going Christian, among whom the proportion of tax-dodgers is much less than 50\%. Other evidence would have to be provided by the State to make a convincing case, e.g. is he living beyond his means? He declared an average of \$15,000 income per year for the past few years, yet he lives in a big house and recently bought several fancy cars, etc.

Second, in the Bayesian philosophy, probability is a subjective measure that applies to individual cases, but it is based on a subjective bet. The subjective probability is the price that you would be happy to pay in order to win \$1.00 if your bet is correct. It is not at all obvious that judges or juries are inclined, or even allowed, to think in that way. Strictly, in a betting perspective, an agent can make a bet based on whatever information available to him, but what guarantees the legal relevance of the bet to the unique case? In the tax-cheating example, suppose you're the judge. \emph{If there is no other information available to you} then the statistical probability 0.60 is the fair price for a bet that Mr.\ John is a cheater. If you're happy with the price then you will bet that he is guilty of tax evasion. How does the subjective betting decision help the cause of justice?

The full decision process by a jury is perhaps too complex to be captured by a single number such as the probability the defendant is guilty. In \emph{The Probable and the Provable}, Cohen (1971) gave examples to indicate why a high probability alone is not a sufficient reason to convict. A probability is something offered by an expert witness, typically for a piece of evidence, not something estimated by a juror as a measure of the quality of his/her decision. Here is an example from Cohen: A has been accused of murdering B, and one piece of evidence is a threatening letter found in B's house. It was not signed, but an expert witness identified three unusual characteristics in the typed letter that could be produced by A's typewriter. 

Suppose the chance of a randomly chosen typewriter could produce the three characteristics together is 1 in a million. So, there is a high likelihood that the letter was typed on A's typewriter. There are still many other questions that the jury will want to know or the defence lawyers will bring up to keep reasonable doubts in the mind of the jury. Did A actually type the letter? Could there be someone else who used the typewriter in trying to frame him? If he did it, did he do it voluntarily? And, anyway, how did the letter establish him as a murderer? Did he have the opportunity to kill B? And if he had, did he take advantage of it? Perhaps he had a strong alibi? So, there is a wide gap between the high probability of one circumstantial evidence and the removal of reasonable doubt.

\subsection{DNA fingerprinting}

DNA profiling, first developed in the mid 1980s by the geneticist Alec Jeffreys, is now commonly used in courts. It is a forensic technique to establish the `genetic fingerprint' of any biological sample -- such as blood, semen or hair -- collected from a person or a crime scene.     DNA profiling has also been used in civil cases such as in parental disputes, but to be specific we shall use the language of a criminal court. There are many technical issues in the statistical analysis of the data (e.g. Roeder, 1994), but to make a long story short, the final evidence is typically presented to the jury in terms of a likelihood ratio. The numerator likelihood is the probability of a match between the DNA profile of the crime-scene sample and the DNA profile of the suspect. It is typically assumed that the test is fully sensitive such that if the suspect is the true source of the sample then the test will produce a positive result. So, the likelihood is
$$
L(\mbox{Suspect is the source})=\P(\mbox{DNA match$|$Suspect is the source}) = 1,
$$
where for convenience the symbol `$|$' can be read as the conditional marker `given that,' even though formally the condition is a fixed hypothesis or a parameter, not a realization of a random event.

The denominator likelihood is the probability of a match if the suspect is not the source. A perfect test should produce a negative result, i.e. the probability of a match is zero. This is now possible with the whole-genome sequencing, but it is still not in common use. Most commonly used DNA profiling is based on a limited number of markers; since 2017 forensic labs in the US used a set of 20 markers  (https://www.nist.gov/news-events/news/2016/12/nist-research-enables-enhanced-dna-fingerprints). Not all of the 20 markers might be readable in a single sample. A person's DNA profile is not fully specific: there could be person(s) in the background population -- where the suspect belongs to -- that has a similar DNA profile as the suspect. This depends on the statistical information about the DNA profile in that relevant population. So the probability of the match is the probability that a random person has a DNA profile that matches the DNA profile of the crime sample. While non zero, this probability is really small; estimating the size is part of the technicalities I refer to above. So, in general
$$
L(\mbox{Suspect is not the source})=\P(\mbox{DNA match$|$Suspect is not the source}) = p,
$$
where $p$ is a tiny number, such as $10^{-9}$ or even much lower. The DNA evidence is then presented as the likelihood ratio
\begin{equation}
R_1 = \frac{L(\mbox{Suspect is the source})}
{L(\mbox{Suspect is not the source})} = 1/p,
\label{eq:R1}
\end{equation}
so $R_1$ is an impressive big number like a billion or more.

With a DNA match, the jury could then be presented with statements like `the suspect is a billion times more likely to be guilty than not guilty.' But as for the typewriter case, the likelihood is actually not the likelihood of guilt directly. The defence would still ask, and the jury would want to know, for example: where is the evidence that the suspect was actually at the crime scene? Otherwise, someone -- God forbids, the police -- might have tried to frame him. Or, the suspect might have a strong alibi that he was somewhere else. If there was a weapon involved, where is it? Is there evidence that the suspect had the weapon prior to the murder? Etc, etc. So, the jury may have reasonable doubts not because the presented probability did not reach certainty, but because there are specific evidence that they still miss. 

\subsection{Database search}
When DNA profiling is performed on a suspect, there is little controversy about the use of the likelihood ratio $R$ to represent the level of evidence against the suspect. But how do assess the evidence if the police have no suspect, but instead trawl through a database containing DNA profiles of many individuals?  The US Federal Bureau of Investigation maintains a database of DNA profiles called CODIS (Combined DNA Index System); by April 2021 it contains $\sim$20M profiles from (i) convicted felons ($>$14M profiles), (ii) persons who had ever been arrested ($>$4M profiles), (iii) forensic samples from crime scenes ($>$1M profiles) and (iv) relatives of missing persons. Since its initiation in 1998, CODIS had helped more than 500,000 investigations. For more details, see {https:// www.fbi.gov/news/pressrel/press-releases/the-fbis-combined-dna-index-system-codis-hits-major-milestone and https://www.fbi.gov/services/laboratory/biometric-analysis/codis}

To see that there is an issue, let's imagine a card trick: a magician shows a standard deck of playing cards and asks you to shuffle them. He takes the deck back face down, chooses one card and puts it aside, still face down so no one can see what it is. He asks you to think of a card then name it out loud. You say `ten of diamonds.' The magician then reveals the card and, sure enough, it is the ten of diamonds. Everyone is duly impressed, as the probability is 1 in 52 for the magician to select the correct card. This is what happens when a hot suspect, who is like the chosen card, matches the crime sample. The small probability of a match explains the magical element of surprise experienced by the spectators. 

If I simply search for the ten of diamonds among the 52 cards and show it, no spectator will be impressed. This is what happens in a cold search. Having no suspect, the police would send the DNA profile from the crime scene to CODIS database for a search. If there is a good match, then the Matcher will be considered a suspect for a closer investigation. Let's call him a cold suspect. Now the question is, is the DNA evidence against the cold suspect from the DNA profile weaker than the evidence against a hot suspect? If we follow the intuition from the card trick, then the answer seems to be yes. If yes, by what factor? In the card trick, the evidence is weakened by 52 fold, so there is no surprise at all. Suppose the search is limited to convicted felons, so there are roughly 14M profiles. Is the evidence weakened by 14M folds? This would negate much of strength in the DNA evidence.

This is where the controversy starts. In line with the reasoning above, an expert committee of the US National Research Council (NRC, 1996) recommended that the likelihood ratio $R$ from a cold search is divided by the size of the database, so the strength of evidence is much reduced. We may call this the frequentist solution. Yet most statisticians on the Bayesian side  claim that the cold search has no negative effect on the strength of DNA evidence. (see for instance the Discussion by Dawid in Lindley, 2000, and Donnelly and Friedman, 1999, which we'll refer to as DF99). In fact they would even go further: the evidence is now stronger than in the hot-suspect case, because the database search has ruled out all other suspects.  Who is right? The pure likelihood-based reasoning indicates that neither is completely right.

\subsubsection*{Frequentist solution}
Let $D+1$ be the number of profiles in the database;  E is the event that there is a single match between the DNA profile of the crime sample and the profiles in the database, and call the person the Matcher; SID be the the hypothesis that the source of the crime sample is in the database (you may think of the source as the criminal but, as we have discussed, legally the guilty status is yet to be decided); SND the source is not in the database. As before, assume that the DNA profiling is fully sensitive, such that
$$
\P(\mbox{E$|$SID}) =1\times (1-p)^D.
$$
Given SND, the event E is a binomial event with probability 
$$
\P(\mbox{E$|$SND}) =(D+1)p(1-p)^{D},
$$
where $p$ is the probability of a DNA match for a random person. A key note: using the binomial probability here means that we're only using the information that there is a single match in the database, even though we actually know the particular Matcher. The likelihood ratio is
\begin{equation}
R_F = \frac{(1-p)^{D}}{(D+1)p(1-p)^{D}}= \frac{R_1}{D+1}\le R_1, \label{eq:NRC}
\end{equation}
where $R_1$ is the likelihood ratio (\ref{eq:R1}) in the hot-suspect case. So the evidence in the cold search is weaker than in the hot search, and that is in line with the NRC recommendation. It also corresponds to our intuition in the magic card trick: there is nothing magical if a match is found after searching the deck of cards. (Note: there is another frequentist solution based on a different probability reasoning that ends up with the same likelihood ratio as an approximation; see DF99.)

\subsubsection*{Bayesian solution}
The main criticism from the Bayesian side, such as DF99, is that $R_F$ represents evidence that is not relevant to the court, that `somebody in the database is the source of the crime sample.' What matters to the court is whether a particular person -- the Matcher -- is the source. According to DF99 the likelihood ratio should be 
\begin{equation}
R_B = \frac{R_1}{\P(\SND)} \ge R_1,\label{eq:DF99}
\end{equation}
where $\P(\SND)$ is the prior probability of the source not in the database before any DNA evidence; for convenience a derivation is given in the Appendix. This prior probability might depend, for instance, on the size, quality and relevance of the database. The likelihood ratio $R_B$ is generally larger the ratio in the hot pursuit, thus representing a stronger evidence. DF99 attributed this feature to the fact that many potential suspects have been ruled out. $R_B$ also depends on the size of the database in a completely opposite way to $R_F$. As the database gets larger $R_F$ gets smaller, but $\P(\SND)$ will naturally drop so $R_B$ gets larger. If the database is so large that it covers all potential suspects then $\P(\SND)=0$, i.e.\ the source is guaranteed to be in the database and the Matcher is guaranteed to be the source. This corresponds to $R_B=\infty$, which makes sense. This scenario has been highlighted by DF99 as  a virtue of $R_B$ and a weakness of $R_F$. 

But what would happen at the other extreme when $\P(\SND)=1$? That is, for whatever reasons, we have an a priori knowledge that the source cannot be in the database; for instance the database is a database of dead criminals or newborn babies. According to (\ref{eq:DF99}) we have $R_B=R_1$, the ratio in the hot-pursuit case. But that does not make sense: in this case we know that any Matcher must be a false positive, so the ratio should be zero.

\subsubsection*{Pure likelihood solution}
Both of the frequentist and Bayesian solutions have some weaknesses. I will now present a pure likelihood solution that overcomes them and highlights the evidential value of the likelihood. What bothers the frequentists in the cold search is that, in contrast to the hot search, we do not specify any hypotheses in advance, i.e., before seeing the match. In terms of the card trick, the magician's pre-selection of the card makes the feat impressive, while trawling through the deck of cards after a card is mentioned looks decidedly unmagical.

But we \emph{can} specify the hypotheses \emph{in advance.} Let $\theta_i$ be the unknown binary status of whether person/profile $i$ is the source of the crime sample, for $i=1,\ldots,D+1$. Let $\theta=(\theta_1,\ldots,\theta_{D+1})$ the vector of the parameters. Define (i) a collection of $(D+1)$ hypotheses: person $i$ is the source for $i=1,\ldots,D+1,$ and (ii) the null hypothesis that the source is not in the database. Formally, in terms of the parameters, the hypotheses are: $(\theta_i=1,\theta_j=0, j\ne i)$ for $i=1,\ldots,D+1,$ and $(\theta_1=0,\ldots,\theta_{D+1}=0)$. Let $y=(y_1,\ldots,y_{D+1})$ be the observed match between the profiles in the database and the crime-sample profile. Assume that there is a single match, and, without loss of generality, let's call this Matcher as person 1, so we have observed the data: $y_1=1$ and $y_i=0$ for $i>1.$ 
The main hypothesis of interest is that person 1 is the one and only source, $\theta_{11}\equiv(\theta_1=1,\theta_2=0,\ldots,\theta_{D+1}=0)$.
The likelihood is
\begin{equation}
L(\theta_{11};y)=P_{\theta_{11}}(y_1=1,y_2=0,\ldots,y_{D+1}=0)=1\times(1-p)^{D}.
\label{eq:like-num}
\end{equation}
Any competing hypotheses with $\theta_i=1$ for $i\ne 1$ have zero likelihood by the assumption of full sensitivity of the test, so they are ruled out. The only other non-zero likelihood is for $\theta_{10}\equiv (\theta_1=0,\ldots,\theta_{D+1}=0)$, so the Matcher is not a source. It is
\begin{equation}
L(\theta_{10};y) = P_{\theta_{10}}(y_1=1,y_2=0,\ldots,y_{D+1}=0)=p\times(1-p)^{D}.
\label{eq:like-denom}
\end{equation}
So we end up with the likelihood ratio
\begin{equation}
R_L = L(\theta_{11};y)/L(\theta_{10};y) = R_1.\label{eq:like}
\end{equation}
for the Matcher being the source vs not the source, exactly the same likelihood ratio as in the hot-pursuit case! Assuming that we want a likelihood ratio that \emph{purely represents the evidence in the data}, the DNA evidence in a cold search is the same as in the hot search. For some people this result might be intuitively obvious, while for others surprising. I will explain more below.

As in DF99, before the DNA search, we may have a prior likelihood ratio that the source of the crime sample is in the database vs not in the database. (Let's put aside the legal issue whether the court would allow a formal prior assessment.) Using the same notation as DF99, the ratio can be written as 
$$
R_0 = \P(\SID)/\P(\SND),
$$
where $\P(\SID)\equiv 1-\P(\SND).$ Since we're not in the Bayesian framework, SND is a fixed hypothesis that does not have a probability and we should interpret $\P(\SND)$ as prior likelihood, not prior probability. Now, if the source is in the database then the Matcher is the source, and vice versa if the source is not in the database the Matcher is not the source. So SID is equivalent to the hypothesis that the Matcher is the source, and SND equivalent to the Matcher not being the source. So the prior likelihood ratio
$$
L_0(\theta_{11})/L_0(\theta_{10}) = \P(\SID)/\P(\SND)=R_0,
$$
and the total likelihood ratio including the prior likelihood is
\begin{equation}
R_T = R_0\times R_1.\label{eq:RT}
\end{equation}
Here are some notable features of the total likelihood ratio: Unlike $R_F$ or $R_B$, in general $R_T$ can be larger or smaller than $R_1$. The formula makes it explicit that the pure contribution of the data to the likelihood is in $R_1$, and there is an independent contribution of prior information in $R_0$. When nothing is assumed known about the database (or the court does not allow prior information), then we have the default $R_0=1$ and $R_T=R_1$. Since $R_0=1$ corresponds to $\P(\SND)=0.5$, the Bayesian solution in this case gives $R_B=2R_1$, which is rather unintuitive. 

As usual, the choice of prior likelihood ratio can be controversial. One might consider, for example, the conviction rate of the people in the database vs the rate in the general population. Extreme cases might also be argued: if the database is so large, e.g. including the whole population, that it is guaranteed to contain the source then $\P(\SND)=0$, so $R_0=\infty$ and $R_T=\infty=R_B$. This makes sense as the Matcher is guaranteed to be the source. At the other extreme, when the database is irrelevant so that $\P(\SND)=1$, we have $R_0=0$ and $R_T=0$, which as we expect since the evidence of a DNA match has no value. But in this case $R_B=R_1$, which does not make sense.

\subsubsection*{The card trick revisited}
How do we interpret the fact that the pure likelihood ratio $R_L$ ignores the database search step? In the card trick example it is equivalent to equating the non-magician and the magician routines. Let's consider another card example. You're shown a deck of cards facing down. \emph{Before} you pick a card at random, you're told that the deck of cards is either $H_0$: a standard deck of 52 distinct cards, or $H_{10}$: a special deck containing 52 cards of tens of diamonds only. Then you pick a card at random and it's a ten of diamonds. Which kind of deck is more likely? You'd of course be suspicious that the deck has been rigged, so think that $H_{10}$ is 52 times more likely than $H_0$.

Now, suppose you're not told \emph{in advance} about the cards, but just pick a random card and see a ten of diamonds. Based on this information, which is more likely: $H_{10}$ or $H_0$? Actually, exactly the same as before: $H_{10}$ is 52 times more likely than $H_0$. This feels disturbing since the hypothesis is specified \emph{after} seeing the data. But that's not an issue. As for the likelihood ratio $R_L$ above, we can specify all the possible decks in advance. So, the evidence in the data for the two scenarios is actually the same. What is different, and this affects our sense of surprise, is our prior expectation. 

Our reaction and full processing uncertainty do include both the prior expectation and the likelihood extracted from the data. The mismatch between the two generates a sense of surprise. It can be created by the magician's skill in handling the cards, which is then rewarded by the audience applause. In the court case, the jury would likely feel the same way.  An evidence of DNA match against a hot suspect would feel like a strong evidence that would impress the jury, since the police must have the skill/ability to find other pieces of evidence against the suspect. But, a mere DNA match from a database will likely not convince the jury. It will only be a starting point for the police to build up the case, as they need to find other pieces of evidence in order to establish the guilt beyond reasonable doubt. 

Let's imagine two films about the same crime: one with the narrative going forward, starting with multiple clues that lead to the suspect, which is further supported with a DNA match. The other film goes backward, starting with a DNA match from a database search that identifies a potential suspect. The police then find multiple clues that establish the person as the suspect for the trial. Do you feel any difference in the strength of the total evidence against the suspect in the two narratives?

\section{Conclusions}
It is of course not surprising that uncertain statements generate logical problems not found in certain ones. The lottery paradox highlights one such problem where we knowingly hold contradictory positions: we rationally accept uncertain statements that have high probability of being true, but we also happily accept the opposite of the consequences of those statements. The DNA forensic evidence story shows the limit of the use of probability-based reasoning in court. Putting aside the issue of differential quantification of evidence by the different schools of inference, the resulting probability/likelihood assessment is not of the guilt status of the suspect. Probability is limited to an assessment of a specific evidence. The full decision process of the jury, which needs to assimilate many more pieces of evidence, is perhaps too complex to be summarized into a numerical probability value.

\section*{Appendix: Derivation of $R_B$}
For convenience I show here the derivation of the likelihood ratio $R_B$ (\ref{eq:DF99}) for the cold search in Donnelly and Friedman (1999). As before let $(D+1)$ be the number of persons (=profiles) in the database; $N$ the number potential suspects in the population outside the database; $m$ is a factor such that a person in the database is $m$-times more probable to be the source relative to a person outside the database; the data $E$ is the recorded event that there is a single match in the database by a \emph{particular person} called the Matcher. 

The numerator of the likelihood ratio is the probability of E given that the Matcher is the source of the crime sample. This is the same as the probability in (\ref{eq:like-num}), which is $(1-p)^{D}$. The denominator is the probability of E given that the Matcher is not the source. This probability is computed as a weighted average
$$
\frac{mD\P(E|S_d) + N\P(E|S_p)}{mD+N},
$$
where `$\P(E|S_d)$ is the probability that the evidence would arise if someone other than the Matcher represented in the database were the source of the crime sample and $\P(E|S_p)$ is the probability that the evidence would arise if someone in the suspect population not represented in the database were the source.' The weights are the prior probabilities that the source is inside or outside the database, respectively, before the search. 

The probability $\P(E|S_d)=0$ as $E$ is impossible under the assumption that the DNA test is fully sensitive. And, if the source is outside the database, 
$$
\P(E|S_p) = p\times(1-p)^{D},
$$
exactly the same meaning and value as the probability (\ref{eq:like-denom}). So, the  likelihood ratio is 
\begin{eqnarray*}
R_B &=& \frac{(1-p)^{D}}{Np\times(1-p)^{D}/(mD+N)}\\
&=&\frac{R_1}{N/(mD+N)} = \frac{R_1}{\P(\SND)},
\end{eqnarray*}
which is the hot-suspect likelihood ration $R_1$ divided by the prior probability that the source is outside the database before the search. As $N\rightarrow 0$ or $m\rightarrow\infty$, the database is guaranteed to contain the suspect, so $\P(\SND)\rightarrow0$ and $R_B\rightarrow \infty$. When $m=0$, the source must be outside the database, so $\P(\SND)=1$, but $R_B=R_1$.

\section*{References}
\begin{description}

\item Cohen, L.J. (1977). \emph{The Probable and the Provable,} Oxford: Clarendon Press.

\item Donnelly, P.\ and Friedman, R.D.\ (1999) DNA database searches and the legal consumption of scientific evidence. \emph{Mich. Law Rev.}, 97, 931-984.

\item Kyburg, H.\ E.\ (1961). \emph{Probability and the Logic of Rational Belief}. Middletown, CT: Wesleyan University Press.

\item Lindley, D.V.\ (2000) The philosophy of statistics. \emph{The Statistician} (2000) 49, pp.\ 293-337.  

\item National Research Council (1996). \emph{The Evaluation of Forensic DNA Evidence.} Washington, DC: The National Academies Press. https://doi.org/10.17226/5141.

\item Roeder, K. (1994). DNA Fingerprinting: A Review of the Controversy. \emph{Statist. Sci.} 9(2): 222-247.
\end{description}

\end{document}